\documentclass[aps,twocolumn,pra]{revtex4-1}

\usepackage{amsmath,amstext,amssymb,graphicx,bm,dcolumn,color,times, braket, multirow}
\usepackage[colorlinks=true,citecolor=blue,urlcolor=red]{hyperref}
\usepackage[table]{xcolor}

\begin{document}

\title{Canonical Leggett-Garg Inequality: Nonclassicality of temporal quantum correlations \\ under energy constraint}

\author{Titas Chanda${}^1$, Tamoghna Das${}^{1,2}$, Shiladitya Mal${}^{1}$, Aditi Sen(De)${}^1$, and Ujjwal Sen${}^1$}

\affiliation{${}^1$Harish-Chandra Research Institute, HBNI, Chhatnag Road, Jhunsi, Allahabad 211019, India \\
${}^2$Institute of Informatics, National Quantum Information Centre, Faculty of Mathematics, Physics and Informatics,
University of Gdan\'{s}k, 80-308 Gdan\'{s}k, Poland
}

\begin{abstract}
Nonclassicality of temporal correlations pertaining to non-commutative sequential measurements is defined through the violation of macrorealistic inequalities, known as Leggett-Garg inequalities (LGI). We investigate the energy cost of the process associated with the Leggett-Garg test in the context of noiseless and Markovian noise for arbitrary initial states. We prove that in the noiseless and certain noisy scenarios, the maximal violations of LGI under the energy constraint occurs when the average energy of the process is equal to the negative of the energy of the initial state. Such a dependence of LGI on the choice of the initial state is not seen in the unconstrained case. Moreover, we find that in presence of a moderate amount of Markovian noise,
the amount of violation of LGI  remains almost unaltered with suitable choice of the evolution and dephasing operators in the neighborhood of the maximal violation line, thereby showing the robustness of temporal correlations under environmental effects.

\end{abstract}

\maketitle

\section{Introduction}

In quantum theoretic description of nature, correlation between measurement outcomes arises in various ways. Compatible measurements performed on spatially separated systems lead to spatial correlation, violating Bell's inequality \cite{bell'65, brunner'2014} 
while sequential measurements  on a single system give rise to temporal correlation. Such correlation pertaining to commuting sequential measurements can not be described by non-contextual model, revealing contextual nature of quantum theory when dimension of the system becomes greater than  or equal to three \cite{kochen, kcbs}. On the other hand, sequential non-commuting measurements on a single quantum system yields temporal correlations which are incompatible with \emph{macroscopic realism} and \emph{noninvasive measurability} \cite{leggett'1985}, leading to the violation of macrorealistic models. 
Based on these assumptions, which are compatible with classical physics, one can derive an inequality, known as Leggett-Garg inequality (LGI), which is violated by nonclassical temporal correlations \cite{leggett'1985, emary'2014}.   
%
%
%

Violation of LGI by quantum mechanical systems was first demonstrated in the laboratory by employing superconducting qubits \cite{lgi_exp_sc} (see  also \cite{van'2000, friedman'2000, roskov'2006, jordan'2006, mahesh'2011, leggett'2008, emary'2014} for other experiments and experimental proposals). Since then,
such studies, which
include generalizations of the original Leggett-Garg (LG) test \cite{leggett'2002, kofler'2007, avis'2010, onofrio'1999, home'2013, budroni'2013, dressel'2014, budroni'2014, mal'2016_1, maldas'2016}, inequivalent necessary conditions for 
macrorealistic theories like no signaling in time \cite{kofler'2013}, entropic LGI 
\cite{udevi'2013}, Wigner's form of LGI \cite{mal'2015} have been carried out. 
Initial motivation of examining LGI was to address a perennial question of probing quantum superposition for macroscopic objects and therefore to relate it with the problem of quantum-classical transition. Apart from the fundamental importance, nonclassicality of temporal correlations has also been employed in various information processing tasks. For example, contextuality of quantum theory has applications in quantum key distribution protocol \cite{cabello, arvind}. Recently it has been shown that sequential transformation contextuality can be probed using single qubit systems and has application in quantum computing of non-linear functions \cite{shane}. Nonclassical temporal correlation revealed through violation of LGI has been employed in quantum computation \cite{brukner'2004}, device independent randomness generation \cite{mal'2016_2} and secure key distribution \cite{shenoy'2017}, thereby showing its significance as important resource.

In reality, every physical process is performed within the realm of limited resources. For example, it is known  that classical capacity \cite{holevo'1973, schumacher'1996} of a noiseless quantum channel \cite{lloyd'2003} can be infinite for infinite-dimensional system, such as for bosonic channels \cite{lloyd'2003,asen'2005,  asen'2007}. 
To overcome such nonphysical situation of infinite capacity, one requires to optimize the capacity under energy constraint.
Moreover,
in infinite-dimensional pure system, the average energy constraint is also related to the continuity property of the entanglement measure \cite{eisert'2002, shirokov'2004}, as quantified by the local von Neumann entropy. Specifically, without the energy constraint, one can find that there exists states having infinite entropy of entanglement which are close to the pure product states with vanishing entanglement.
One can circumvent such discontinuity by imposing additional constraints like bounded mean energy of the state \cite{eisert'2002}. Such bridge between correlation and energy have also recently been explored in finite dimensional systems \cite{thcost1, thcos, horo'1997, horo'1998, horo'98, horo'01, tamoghna'2017}.

The present work aims to establish a connection between the thermodynamics, via energy considerations, and macrorealism, by employing LGI -- we call it as ``canonical LGI".
 Specifically, we ask the following question: For a given amount of energy available in the system, how much violation of LGI can be obtained by choosing suitable measurement strategies and dynamics? 
To this end, we find that although LGI is independent of initial state, the maximal violation of LGI for a given energy cost depends upon the initial state in noiseless as well as in noisy scenarios.
In particular, we prove that maximal violation of three-termed LGI occurs if and only if the energy cost is  the negative of the 
average energy of the initial state. Motivated by practical situations, we then consider scenarios where evolution is affected by dephasing Markovian noise described by Lindblad master equations. 
In this situation, we consider a few paradigmatic models with arbitrary pure and mixed states as initial ones -- (i) the dephasing operator and the evolving Hamiltonian are diagonal in the same basis, (ii) they are in complementary bases, and (iii) they  are in bases which are at an angle $\pi/4$ to each other. 
When the evolution starts from an arbitrary single qubit pure  state, we obtain the following results: 
In case (i), the violation decreases monotonically with noise parameter without 
showing any decrement/shrinking in the violating region compared to the noiseless case in the parameter space of the initial state and the energy cost.
In contrast, in the second situation, the system in terms of violation of LGI shows much more robustness against  noise than the first case, although the region of violation reduces with the increase of noise. Just like in the noiseless scenario, we prove that in both the cases (i) and (ii), maximum violation of LGI occurs when the change in energy is equal to the negative of the average energy of the initial state. The third scenario is the intermediate case of (i) and (ii). Interestingly however, here, the maximum violation of LGI gets shifted with increasing noise parameter, which is not seen in situations (i) and (ii). 
This is true in general when the Hamiltonian and the dephasing operator are not diagonal in the same basis or in complementary bases. 
Qualitatively similar characteristics are observed for mixed states.

The paper is organized in the following way. In Sec. \ref{sec:def}, we discuss the formalism of calculating energy cost for an LG test. In Sec. \ref{sec:coh_dyn}, we provide the condition on energy for obtaining maximal violation of LGI in the noiseless scenario. The effects of dephasing on energy cost and violation of LGI are discussed in Sec. \ref{sec:dephase_dyn}. We end with concluding comments in Sec. \ref{sec:conclusion}.   

\begin{figure}
\includegraphics[width = \linewidth]{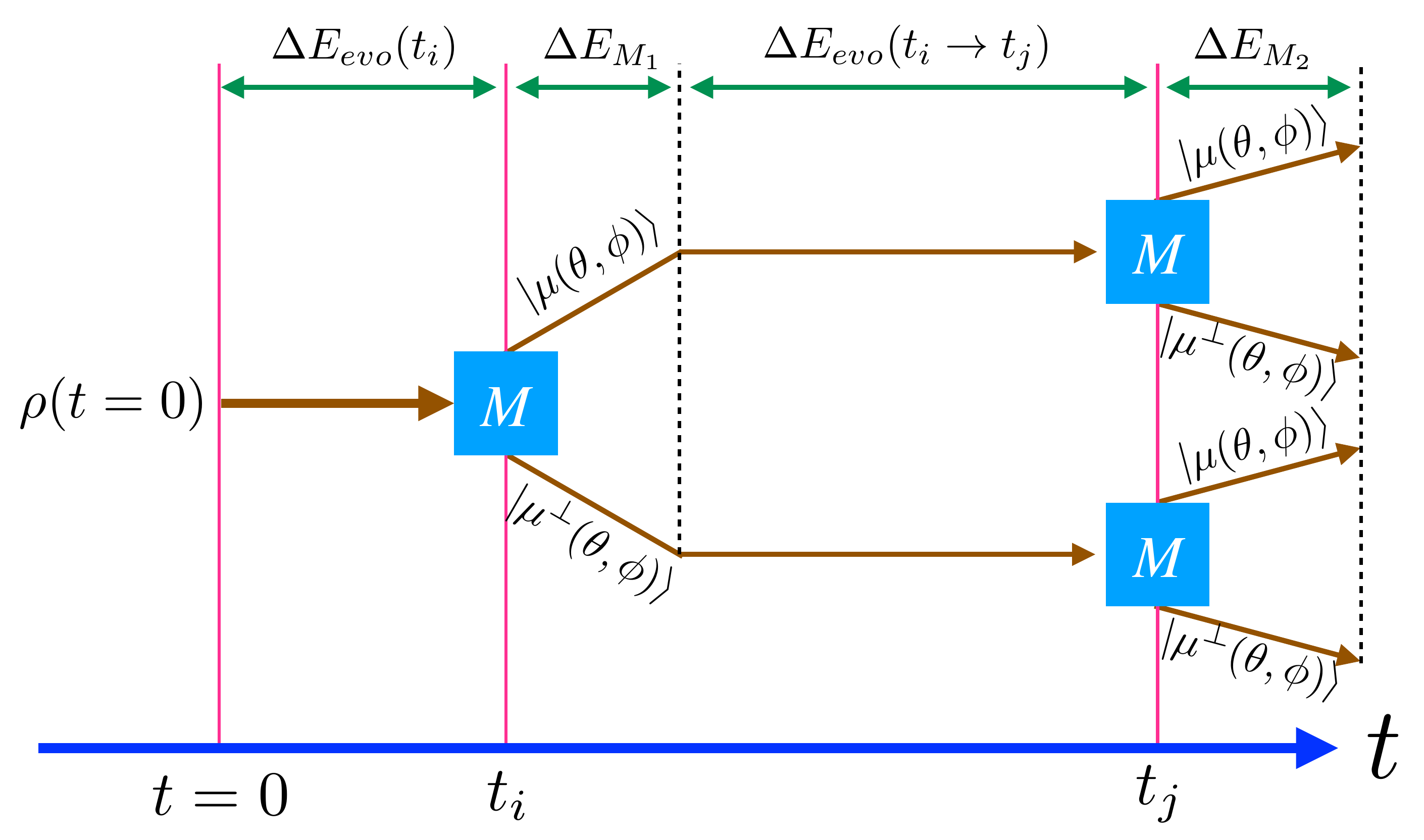}
\caption{(Color online.) Schematic diagram for obtaining the energy change or energy cost, denoted by $\Delta E_{ij}$, corresponding to the correlation function $C_{ij}$.}
\label{fig:schematic}
\end{figure}

\begin{figure}
\includegraphics[width=0.8\linewidth]{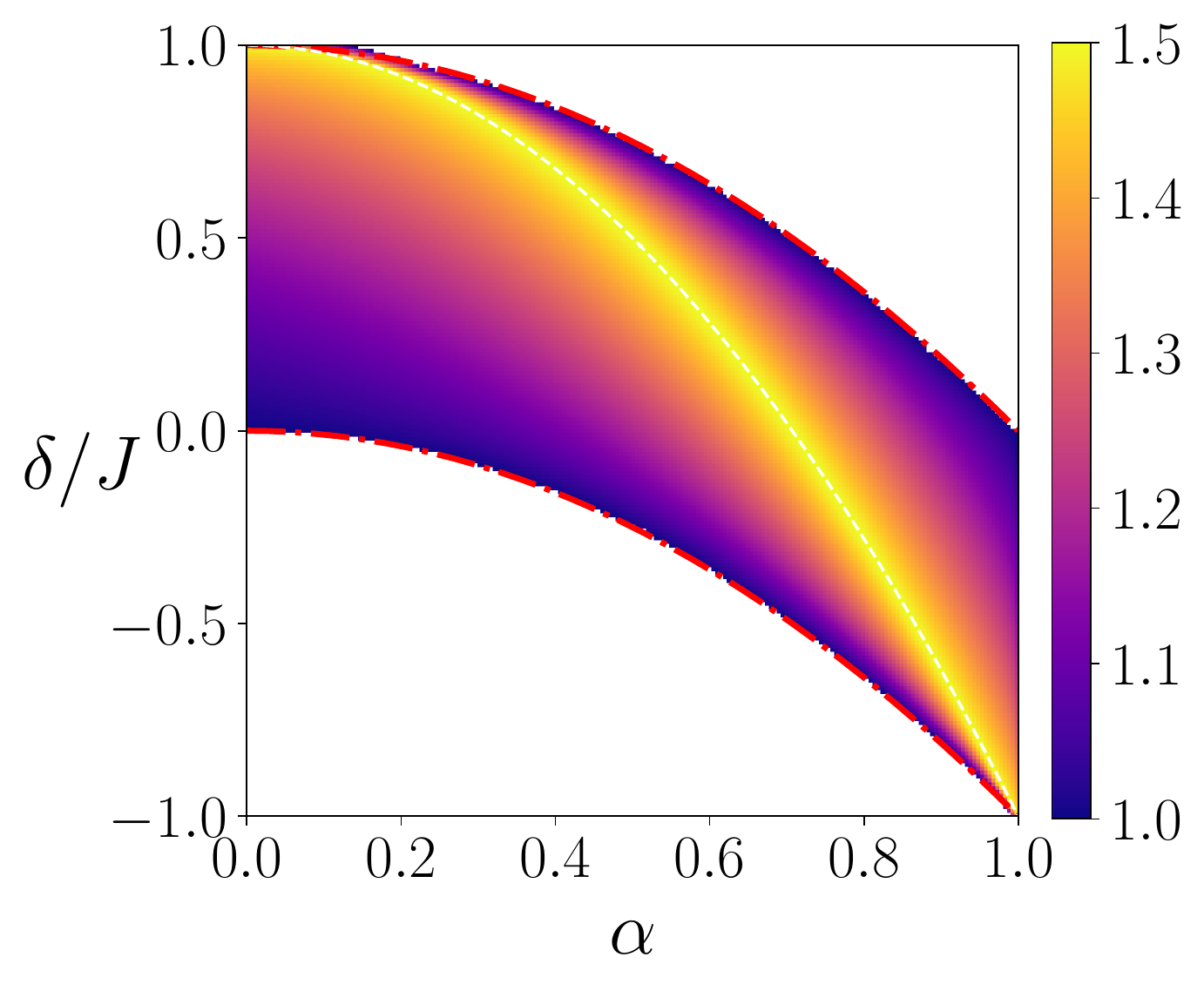}
\caption{(Color online.) Violation of canonical LGI, $K_3^{\mbox{opt}}$, for the initial pure states, $\ket{\psi(\alpha)} = \alpha \ket{0} + \sqrt{1-\alpha^2}\ket{1}$, as a function of $\alpha$ and $\delta /J$.
This is a noiseless scenario.
The region bounded by the (red) dash-dot lines is physically accessible one, where solutions for energy constraint can exist. Yellow dashed line shows the maximum value of $K_3^{\mbox{opt}}$. All quantities plotted are dimensionless.
}
\label{fig:pure_coh}
\end{figure}

\section{Definitions and formalism}
\label{sec:def}

According to the postulate of
quantum mechanics, a system, initially in a pure or mixed state, evolves with time under certain Hamiltonian. We consider series of measurements performed on the same initial state evolving under that Hamiltonian in such a way that in the first series, a dichotomic observable $Q$, which can take values $+1$ or $-1$, is measured at times $ t_{1}$ and $t_{2}$, in the second at $t_{2}$ and $t_{3}$, in the third at $t_{1}$ and $t_{3}$ (with $ t_{1} < t_{2} < t_{3}$). 
From such series of measurements, one obtains joint probabilities of obtaining
the outcome, $q_i$ at time $t_i$ and 
 $q_j$  at time $t_j$, denoted by $P(q_i, q_j)$. The two-time correlation function is then defined as 
\begin{eqnarray}
C_{ij} = \sum_{q_i, q_j = \pm 1} q_i q_j P(q_i, q_j),
\end{eqnarray}
which can be evaluated from the 
joint probability $P(q_i, q_j)$. The three-termed Leggett-Garg inequality then reads as \cite{leggett'1985}
\begin{eqnarray}
K_3 \equiv C_{12} + C_{23} - C_{13} \leq 1,
\label{eq:k3_def}
\end{eqnarray}
which is derived with the assumptions of macroscopic realism and noninvasive measurability.
Let us discuss briefly these two assumptions -- 
{\it macroscopic realism} (MR): at any instant, irrespective of measurement, a system is in any one of the available definite states such that all its observable properties have definite values; {\it noninvasive measurability} (NIM) states that it is possible, in principle, to determine which of the states the system is in, without affecting the state itself or the system's subsequent evolution. 
It can be shown that quantum mechanical systems are incompatible with macrorealistic theories, consisting of MR and NIM, and can violate the $K_3$ inequality mentioned in (\ref{eq:k3_def}). 
For example, 
a two-level system  violates the inequality and the  
 maximum achievable value of the $K_3$-quantity in two dimension is $1.5$ (for detailed review, see \cite{emary'2014}).

It is noteworthy to mention here that the general form of the usual LGI involving $n$ pairs of two-time correlation functions can be expressed as \cite{emary'2014} 
\begin{eqnarray}
&& - n\leq K_n \leq n-2 ~~ \text{for odd $n\geq 3$,} \nonumber \\
&& -(n-2) \leq K_n\leq n-2 ~~ \text{for even $n\geq 4$,}
\label{eq13}
\end{eqnarray}
where $K_n = C_{12} + C_{23} + C_{34} + ..... + C_{(n-1)n} - C_{1n}$.  
It has been shown that
the maximum quantum mechanical value of $K_n$ for projective measurements can be $n \cos(\pi/n)$ \cite{emary'2014}. In the subsequent discussions, we adhere to the three-termed LGI (Eq. (\ref{eq:k3_def})).

Let us investigate the energetics associated with the process contingent to temporal correlations involved in $K_3$. In an experimental scenario, 
the method of obtaining correlation functions inevitably change the energy of the underlying quantum system due to sequential measurements and evolution. For example, in case of acquiring $C_{ij}$, the change in energy, $\Delta E_{ij}$ 
occurs in the following steps (see Fig. \ref{fig:schematic} for the schematic diagram):
\begin{enumerate}
\item The change in energy during the evolution from $t = 0$ to $t = t_i$ is
$\Delta E_{evo}(t_i)$.
\item Measurement at $t_i$ costs energy, hence the difference in energy before and after the measurements is given by $\Delta E_{M_1}$.
\item $\Delta E_{evo}(t_i \rightarrow t_j)$ denotes the change of energy due to the 
 evolution from $t_i$ to $t_j$.
\item  Like step 2, 
$\Delta E_{M_2}$ denotes the energy cost for measurement at $t_j$.
\end{enumerate}
Therefore,
the average change in energy (or, the energy cost) involved in the entire process of finding the Leggett-Garg expression, $K_3$,
can be represented as $\Delta E  = (\Delta E_{12} + \Delta E_{23} + \Delta E_{13})/3$, 
where each $\Delta E_{ij} = \Delta E_{evo}(t_i) + \sum_{i = 1,2} \Delta E_{M_i} + \Delta E_{evo}(t_i \rightarrow t_j)$. In the subsequent analysis,
without loss of generality,
we consider $t_1 = 0$ and $t_2 - t_1 = t_3 - t_2 = \Delta t$.  
In this paper, we are interested in the optimal amount of violation of LGI, i.e., the maximal value of $K_3$ for a given energy cost, $\Delta E$. For this, we restrict our analysis to two-level systems only.
Since we take sequential measurements involved in the process as arbitrary,  
we choose, without loss of generality, an underlying Hamiltonian as ${H} = J {\sigma}_z$, responsible for the time evolution of the system, with $J$ being the unit of energy, and $\sigma_i \ (i = x, y, \mbox{ or } z)$ is the Pauli Matrix.
Note that the energy considered here is the 
the average energy of any state $\rho$ with respect to the underlying  Hamiltonian $H$, given by $\langle E \rangle_{\rho} = \mbox{tr}(H \rho)$. 
Moreover
measurement of a dichotomic observable, $\{Q_{\theta, \phi}\}$, is equivalent to measurement of spin component along a direction determined by $\theta$ and $\phi$, and hence
\begin{eqnarray}
Q_{\theta, \phi} = \ket{\mu(\theta, \phi)}\bra{\mu(\theta, \phi)} - \ket{\mu^{\perp}(\theta, \phi)}\bra{\mu^{\perp}(\theta, \phi)},  
\end{eqnarray}
where
\begin{eqnarray}
\ket{\mu(\theta, \phi)} &=& \cos\frac{\theta}{2}\ket{0} + e^{i \phi}\sin\frac{\theta}{2}\ket{1}, \nonumber \\
\ket{\mu^{\perp}(\theta, \phi)} &=& \sin\frac{\theta}{2}\ket{0} - e^{i \phi}\cos\frac{\theta}{2}\ket{1}.
\end{eqnarray}
Here, $\theta \in [0,\pi)$, $\phi \in [0, 2 \pi)$, and $\{\ket{0}, \ket{1}\}$ are the eigenstates of the Hamiltonian with energies $J$ and $-J$ respectively. 

\begin{figure*}
    \includegraphics[width=\linewidth]{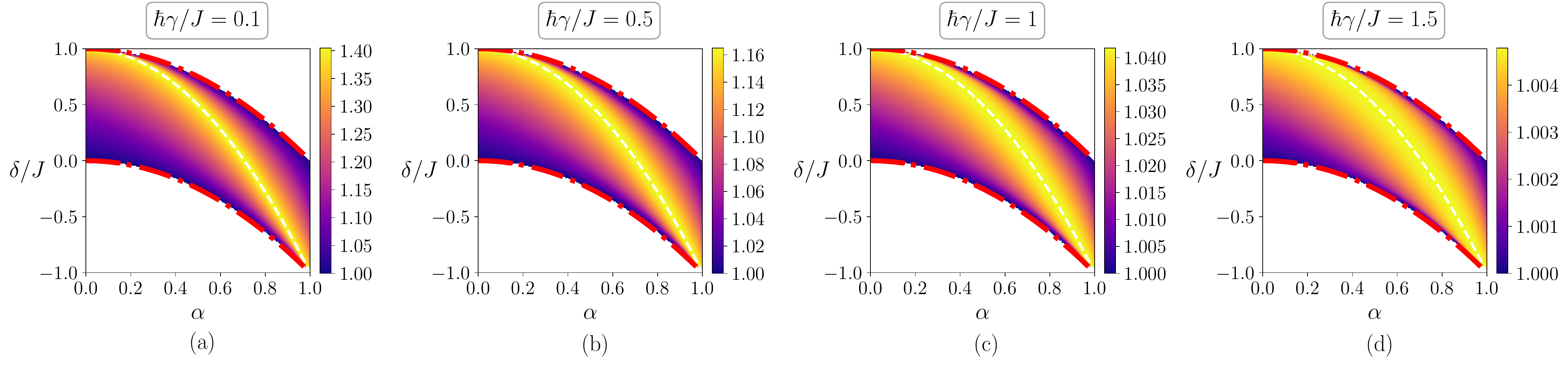}
    \caption{(Color online.) The plot of canonical LG expression, $K_3^{\mbox{opt}}$, for the initial pure states, $\ket{\psi(\alpha)} = \alpha \ket{0} + \sqrt{1-\alpha^2}\ket{1}$, as a function of $\alpha$ and $\delta /J$. The Hamiltonian and the dephasing operator are diagonal in the same basis. Other descriptions are same as in Fig. \ref{fig:pure_coh}. All quantities plotted are dimensionless.}
     \label{puregz}
  \end{figure*}

\begin{figure*}
    \includegraphics[width=\linewidth]{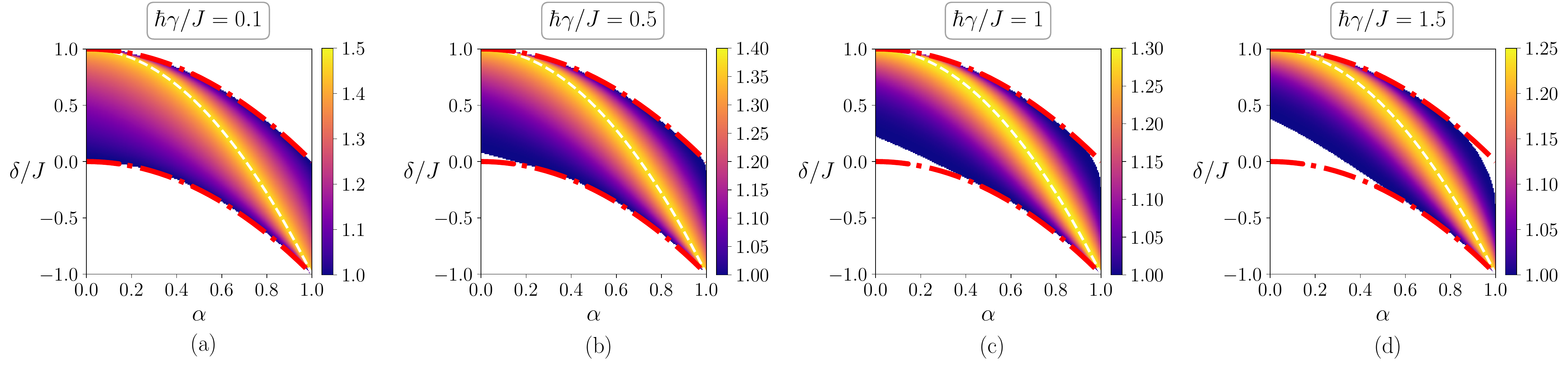}
    \caption{(Color online.) $K_3^{\mbox{opt}}$ against $\alpha$ and $\delta/J$ for the same initial pure states as in Fig. \ref{puregz}. Here the Hamiltonian and the dephasing operator are diagonal in the complementary bases. Other descriptions are same as in Fig. \ref{fig:pure_coh}. All quantities plotted are dimensionless.}
     \label{puregx}
  \end{figure*}

\section{Pure coherent dynamics}
\label{sec:coh_dyn}

Let us first consider a coherent dynamics, i.e., the evolution is without any environmental noise.
Such dynamics can be described  by the unitary operator, $U(t) = \exp(-iHt / \hbar) = \exp(-i J \sigma_z t / \hbar)$. The LG expression under this setting takes the form as
\begin{eqnarray}\label{k3}
	K_3 = 1-2\sin^2\theta \big[2 \sin^2( J \Delta t / \hbar) - \sin^2(2 J \Delta t / \hbar)\big],
	\end{eqnarray}
which is independent of the initial quantum state, $\rho$, and the azimuthal angle $\phi$.
We find that the maximum of $K_3$, 1.5, for the qubit system occurs when the measurement is done in the complimentary basis to the eigenbasis of the Hamiltonian, i.e., for $\theta = \pi/2$ and $\phi$ arbitrary.
The changes in energy for obtaining the correlation functions $C_{12}$, $C_{23}$, and $C_{13}$ in $K_3$ are respectively given by
\footnotesize
\begin{eqnarray}
\Delta E_{12} &=& (p_1 - p_2) (\cos^2\theta + \sin^2\theta\cos(2 J \Delta t / \hbar))\cos\theta - \mbox{tr}(H \rho), \nonumber \\
\Delta E_{23} &=& (p'_1 - p'_2) (\cos^2\theta + \sin^2\theta\cos(2 J \Delta t / \hbar))\cos\theta - \mbox{tr}(H \rho), \nonumber \\
\Delta E_{13} &=& (p_1 - p_2) (\cos^2\theta + \sin^2\theta\cos(4 J \Delta t/ \hbar))\cos\theta - \mbox{tr}(H \rho),
\label{eq:pure_E}
\end{eqnarray}
\normalsize
where
\small
\begin{eqnarray}
p_1 &=& \bra{\mu(\theta, \phi)}\rho\ket{\mu(\theta, \phi)}, \ p_2 = \bra{\mu^{\perp}(\theta, \phi)}\rho\ket{\mu^{\perp}(\theta, \phi)}, \nonumber \\
p'_1 &=& \bra{\mu(\theta, \phi)}U(\Delta t) \rho U^{\dagger}(\Delta t) \ket{\mu(\theta, \phi)}, \nonumber \\
p'_2 &=& \bra{\mu^{\perp}(\theta, \phi)} U(\Delta t) \rho U^{\dagger}(\Delta t)\ket{\mu^{\perp}(\theta, \phi)},
\end{eqnarray}
\normalsize
with $\rho$ being the initial state. 
As mentioned earlier, 
our aim is to find the optimal violation of LGI for a given energy cost.
Therefore, we evaluate the maximum of $K_3$, denoted by $K_3^{\mbox{opt}}$, where the maximization is  over measurement parameters and time interval, $\Delta t$, for a given change in the energy, $\Delta E = \delta$, with $\delta$ being a fixed value, which is chosen according to the relevance of the problem. Hence optimization problem can be represented as 
\begin{eqnarray}
K_3^{\mbox{opt}} = \underset{\theta, \phi, \Delta t} {\mbox{max}} \ K_3, \nonumber \\ \mbox{s.t.} \  \Delta E(\theta, \phi, \Delta t) = \delta.
\end{eqnarray}
We call $K_3^{\mbox{opt}}$ as canonical LG quantity.
Before presenting the results for arbitrary states, let us calculate $K_3^{\mbox{opt}}$ for arbitrary pure states of the form,
\begin{eqnarray}
\ket{\psi(\alpha)} = \alpha \ket{0} + \sqrt{1-\alpha^2}\ket{1},
\label{eq:pure_state}
\end{eqnarray}
where $\alpha \in [0, 1]$. For a particular pure state and for a given energy cost, we find the measurement setting which leads to the maximal violation of LGI. In Fig. \ref{fig:pure_coh}, we plot $K_3^{\mbox{opt}}$ as a function of given energy cost
 $\delta$ and the state-parameter $\alpha$, when evolution is governed by the Hamiltonian $H= J \sigma_z$. In this scenario, we get solutions for the energy-constraint, $\Delta E(\theta, \phi, \Delta t) = \delta$, only if 
\begin{eqnarray}
\delta /J &\leq& 1 - \alpha^2, \nonumber \\
\delta / J &\geq& -\alpha^2.
\end{eqnarray}
 In Fig. \ref{fig:pure_coh},  the region bounded by the (red) dash-dot lines is physically accessible in the
 $(\alpha, \delta/J)$-plane.
 For a given initial pure state $\ket{\psi(\alpha)}$, $K_3^{\mbox{opt}}$ attains its maximum value, 1.5, when
 \begin{eqnarray}
 \delta / J = 1 - 2 \alpha^2,
 \end{eqnarray}
which corresponds to 
$\delta = - \bra{\psi(\alpha)} H \ket{\psi(\alpha)}$, and
 is depicted in the figure by the (yellow) dashed line. 
Interestingly, we notice here that although LGI, in general, does not depend on the initial state, the violation of LGI under energy constraint does depend on the choice of the initial state.  
 With this observation in hand, let us now state the theorem for arbitrary states in two dimension.

\noindent\textbf{Theorem 1.} \emph{In a noiseless scenario, i.e., under pure coherent dynamics,  violation of canonical Leggett-Garg inequality reaches its maximum possible value
 if and only if the change in energy, $\delta$, coincides with the negative value of the initial average energy of the underlying state, $\rho$, i.e., $K_3^{\mbox{opt}} = 1.5$ if and only if $\delta = - \mbox{tr}(\rho H)$.}

\noindent Proof. \\
We first observe that the $K_3$, given by Eq. (\ref{k3}), attains its maximum possible value, $1.5$, when $\theta = \pi/2$, $\phi$ arbitrary, and $\Delta t = \tilde{t} \approx 1.9635$. Clearly, from Eqs. (\ref{eq:pure_E}), we get that $\Delta E$ reduces to $-\mbox{tr}(\rho H)$, when $\theta = \pi/2$.

Conversely, given $\Delta E = \delta = -\mbox{tr}(\rho H)$, we have to optimize $K_3$ with respect to $\theta, \phi, \Delta t$. We notice that one of 
the possible solutions of $\Delta E(\theta, \phi, \Delta t) = \delta= -\mbox{tr}(\rho H)$ occurs for $\theta = \pi/2$. For $\theta =\pi/2$, it is evident from Eq. (\ref{k3}) that
$K_3^{\mbox{opt}} = \underset{\phi, \Delta t}{\mbox{max}} \ K_3 = 1.5$.
\hfill $\blacksquare$

\noindent Therefore, the maximal violation of canonical LGI is dependent on the choice of the initial state, both pure and mixed. We will see that such dependency still remains in presence of noise in the system.
We now state one immediate corollary of the above theorem.

\noindent \textbf{Corollary 1.} \emph{For the well-known states, $\ket{\pm} = (\ket{0} \pm \ket{1})/\sqrt{2}$, $\ket{\pm_y} = (\ket{0} \pm i \ket{1})/\sqrt{2}$, and $\rho = \mathbb{I}_2/2$, there is no energy cost for obtaining the maximal violation of the Leggett-Garg inequality.
}  

\section{Dynamics with dephasing noise}
\label{sec:dephase_dyn}

In this section, we consider situations when the initial (qubit) state undergoes a evolution affected by Markovian dephasing noise. The evolution can no more be described by the unitary operator as discussed in the preceding section. In this case, the evolution of the system is governed by a Lindblad master equation, given by
\begin{eqnarray}
\frac{d\rho}{d t}=-\frac{i}{\hslash}[H,\rho]+V\rho V^{\dagger}-\frac{1}{2}(V^{\dagger}V\rho +\rho VV^{\dagger}),
\end{eqnarray}
where $H = J \sigma_z$, as in the preceding section and $V$ is the dephasing operator, responsible for decoherence.
With this kind of evolution, we now study the violation of LGI for a fixed value of the energy cost obtained in the process of the evaluation of the LG expression.
In this situation, we choose $V$ in three different ways -- 
(1) when the Hamiltonian and the dephasing noise are diagonal in the same basis, (2) when $H$ and $V$ are diagonal in complementary bases, and (3) when $H$ and $V$ are diagonal in bases which are at an angle $\pi/4$ to each other.

\begin{figure*}
    \includegraphics[width=\linewidth]{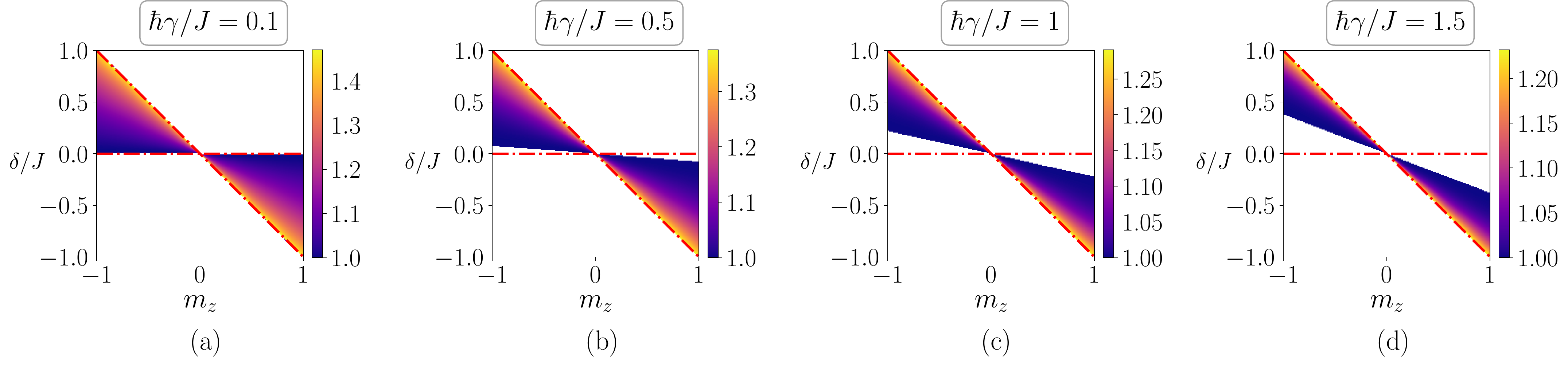}
 \caption{(Color online.) Plot of $K_3^{\mbox{opt}}$, for the initial states $\frac{1}{2}(\mathbb{I} + m_z \sigma_z)$ with respect to $m_z$ and $\delta /J$. Other descriptions are same as Fig. \ref{puregx}. 
All quantities plotted are dimensionless.
}
 \label{mixedzgx}
  \end{figure*}

\subsection{Hamiltonian and dephasing operator are diagonal in same basis}
\label{sec:same_basis}

Let us consider the first case, where both the Hamiltonian  and the dephasing operator are diagonal in the same basis, i.e., 
\begin{eqnarray}
H=J \sigma_z, ~~~~~~~~~~~~~~~~~~~~ V=\sqrt{\frac{\gamma}{2}}\sigma_z.
\end{eqnarray}
Here $\gamma \geq 0$ is the dephasing parameter which controls the amount of noise acting on the system, and is in the unit of $[\mbox{time}^{-1}]$.
Let us take the initial state as
\begin{eqnarray}
\rho(0)=\frac{1}{2}(\mathbb{I}+ m_x(0)\sigma_x +m_y(0)\sigma_y +m_z(0)\sigma_z),
\label{eq:mixed_gen}
\end{eqnarray}
with $m_i \in [0, 1], \ i = x, y, z$. Note that $m_i$'s can take only those values for which $\rho$ remains positive.
After time $t$, it evolves to a state with magnetizations given by
\begin{eqnarray}
	m_x(t) &=& e^{-\gamma t} \big[ m_x(0) \cos(2\frac{J}{\hbar} t) - m_y(0) \sin(2 \frac{J}{\hbar}t) \big], \nonumber\\
	m_y(t) &=& e^{-\gamma t} \big[ m_y(0) \cos(2\frac{J}{\hbar} t) + m_x(0) \sin(2 \frac{J}{\hbar}t) \big], \nonumber \\
	m_z(t) &=& m_z(0).
\end{eqnarray}
The LG expression, $K_3$, for such evolution, then reads as
\begin{eqnarray}
K_3=\cos^2(\theta)+
e^{-2\gamma \Delta t}
 (2 e^{\gamma \Delta t}\cos(J \Delta t/\hbar) \nonumber \\
  -\cos(2 J \Delta t/\hbar))\sin^2(\theta),
 \label{eq:k3_gz}
\end{eqnarray}
\normalsize
which is independent of the initial state $\rho$. For a given decoherence parameter $\gamma$, we again study the optimal violation of LGI, as measured by the canonical LG quantity $K_3^{\mbox{opt}}$, for the energy constraint, $\Delta E(\theta, \phi, \Delta t) = \delta$. Similar to the Theorem 1, we have the following theorem for this noisy scenario.

\noindent\textbf{Theorem 2.} \emph{Violation of canonical Leggett-Garg inequality attains its maximum possible value
 if and only if the change in energy, $\delta$, is equal to the negative of the average energy of the initial state, when the Hamiltonian and the dephasing operator are diagonal in the same basis.}

\noindent Proof. \\
First of all, 
we notice that, 
for a fixed  $\gamma$,
$K_3$, given in Eq. (\ref{eq:k3_gz}),
attains its maximum possible value  when $\theta = \pi/2$ and $\Delta t = \tilde{t}$, where $\tilde{t}$ is the solution of 
$\frac{\partial K_3}{\partial \Delta t} = 0$. The changes in the energy for obtaining $C_{12}$, $C_{23}$, and $C_{13}$ in this case are given by
\footnotesize
\begin{eqnarray}
\Delta E_{12} &=& (p_1 - p_2) (\cos^2\theta + e^{-\Delta t \gamma}\sin^2\theta\cos(\frac{2 J \Delta t}{\hbar}))\cos\theta - \mbox{tr}(H \rho), \nonumber \\
\Delta E_{23} &=& (p_1' - p_2') (\cos^2\theta + e^{-\Delta t \gamma}\sin^2\theta\cos(\frac{2 J \Delta t}{\hbar}))\cos\theta - \mbox{tr}(H \rho), \nonumber \\
\Delta E_{13} &=& (p_1 - p_2) (\cos^2\theta + e^{-2 \Delta t \gamma}\sin^2\theta\cos(\frac{4 J \Delta t}{\hbar}))\cos\theta - \mbox{tr}(H \rho), \nonumber \\
\end{eqnarray}
\normalsize
where
\small
\begin{eqnarray}
p_1 &=& \bra{\mu(\theta, \phi)}\rho(0)\ket{\mu(\theta, \phi)}, \ p_2 = \bra{\mu^{\perp}(\theta, \phi)}\rho(0)\ket{\mu^{\perp}(\theta, \phi)}, \nonumber \\
p_1' &=& \bra{\mu(\theta, \phi)}\rho(\Delta t)\ket{\mu(\theta, \phi)}, \ p_2' = \bra{\mu^{\perp}(\theta, \phi)}\rho(\Delta t)\ket{\mu^{\perp}(\theta, \phi)}. \nonumber \\
\end{eqnarray}
\normalsize
Clearly, $\Delta E$ reduces to $-\mbox{tr}(\rho H)$, when $\theta = \pi/2$.

Now to show the `if' part,  we have to maximize $K_3$ with respect to $\theta, \phi, \Delta t$,
for a fixed $\Delta E = \delta = -\mbox{tr}(\rho H)$. 
As before, one of the possible solutions of $\Delta E(\theta, \phi, \Delta t) = \delta= -\mbox{tr}(\rho H)$ occurs when $\theta = \pi/2$. For $\theta =\pi/2$, $K_3$ indeed reaches its maximum value for $\Delta t = \tilde{t}$.
\hfill $\blacksquare$


In Fig. \ref{puregz}, we map the quantity $K_3^{\mbox{opt}}$ in the  $(\alpha, \delta/J)$-plane for pure states, $\ket{\psi(\alpha)} = \alpha \ket{0} + \sqrt{1-\alpha^2}\ket{1}$, for four different values of $\hbar \gamma /J$, namely 0.1, 0.5, 1.0, and 1.5.
From the figure, we obtain the following observations:
\begin{enumerate}
\item The amount of violation of canonical LGI decreases rapidly with the increase of noise parameter $\gamma$.
\item The region of violation of  LGI in the $(\alpha, \delta / J)$-plane remains same for different values $\gamma$.
This implies that the rate of decrease of $K_3^{\mbox{opt}}$ with $\gamma$ is not uniform for given $\Delta E = \delta$. Specifically, the rate of change of $K_3^{\mbox{opt}}$ with $\gamma$ is faster for high values of $K_3^{\mbox{opt}}$ in the noiseless case, compared to the region having low amount of violation of canonical LGI with $\gamma = 0$.

\item As $\gamma$ increases, dependence of $\delta$ on initial state fades out, although maximum value of $K_3^{\mbox{opt}}$ still occurs for $\delta = - \mbox{tr}(\rho H)$, as stated in Theorem 2.
\end{enumerate}
The situation remains unaltered if one considers mixed states as the initial one, instead of the pure state, i.e., $K_3^{\mbox{opt}}$ decreases rapidly with the noise parameter while the region of violation does not change.
%

\begin{figure*}
    \includegraphics[width=\linewidth]{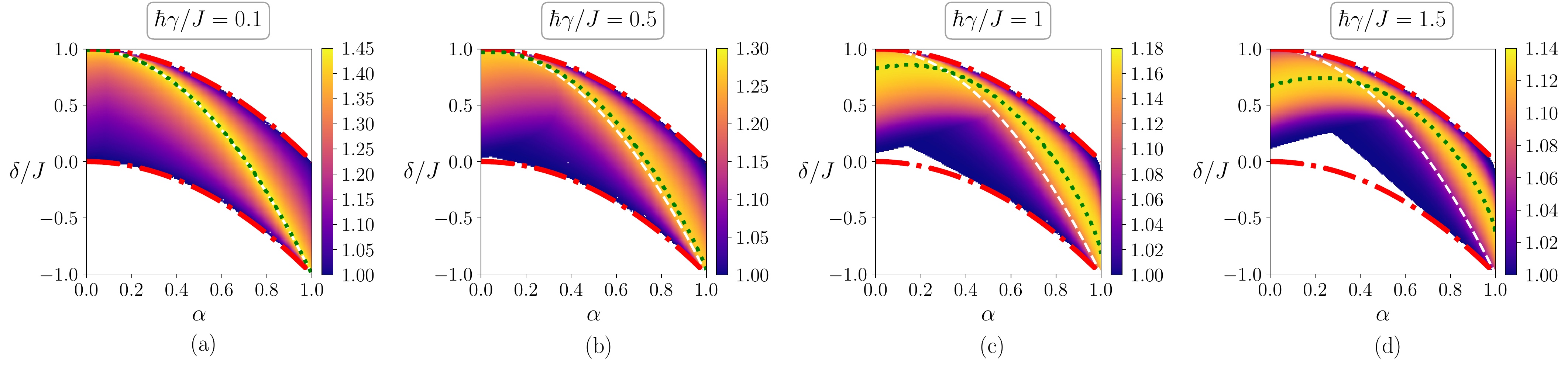}
    \caption{(Color online.) Canonical LG quantity, $K_3^{\mbox{opt}}$, as a function of $\alpha$ and $\delta/J$ for the same initial pure states as in Fig. \ref{puregz}. The Hamiltonian and the dephasing operator are diagonal in bases which are at an angle of $\pi/4$ to each other. 
    Green dotted line shows the maximum value of $K_3^{\mbox{opt}}$, while
    the equation, $\delta = -\mbox{tr}(\rho H)$, is depicted by the white dashed line.
    Other descriptions are same as in Fig. \ref{fig:pure_coh}. All quantities plotted are dimensionless.}
     \label{puregxz}
  \end{figure*}

\begin{table*}[ht]
\begin{tabular}{|c|c|c|c|c|p{2.6cm}|p{5.5cm}|}
\hline
S. No. & $H$ & $V$ & Initial state & Physically accessible region & Maximum violation line ($\delta = - \mbox{tr}(\rho H)$) & Observations \\
\hline \hline 

1 & 
$J \sigma_z$ & $\sqrt{\frac{\gamma}{2}}\sigma_z$ & $\alpha\ket{0} + \sqrt{1-\alpha^2}\ket{1}$ & 
{$\delta/J \leq 1 - \alpha^2$; \ $\delta/J \geq  - \alpha^2$} &
$ \delta /J = 1 - 2\alpha^2$ &
1. $K_3^{\mbox{opt}}$ decreases with increasing $\gamma$. 

2. The region of violation does not change with $\gamma$.
\\

\hline

2 &
$J \sigma_z$ & $\sqrt{\frac{\gamma}{2}}\sigma_x$ & $\alpha\ket{0} + \sqrt{1-\alpha^2}\ket{1}$ & 
{$\delta/J \leq 1 - \alpha^2$; \ $\delta/J \geq  - \alpha^2$} &
$\delta /J = 1 - 2\alpha^2$ &
1. $K_3^{\mbox{opt}}$ remains quite high near the $\delta = - \mbox{tr}(\rho H)$ line.

2. Region of violation shrinks with higher values of $\gamma$.
\\

\hline
3 &
$J \sigma_z$ & $\sqrt{\frac{\gamma}{2}}\sigma_z$ & $(\mathbb{I} + m_x \sigma_x)/2$ & 
{$|\delta/J| \leq |m_x|/2$} &
$ \delta /J = 0$ &
1. $K_3^{\mbox{opt}}$ decreases with increasing $\gamma$. 

2. The region of violation remains same with the variation of $\gamma$.
\\

\hline
4 &
$J \sigma_z$ & $\sqrt{\frac{\gamma}{2}}\sigma_x$ & $(\mathbb{I} + m_x \sigma_x)/2$ & 
{$|\delta/J| \leq |m_x|/2$} &
$ \delta /J = 0$ &
1. $K_3^{\mbox{opt}}$ remains quite high near the $\delta = - \mbox{tr}(\rho H)$ line.

2. The region of violation  does not change with $\gamma$.
\\

\hline

5 &
$J \sigma_z$ & $\sqrt{\frac{\gamma}{2}}\sigma_z$ & $(\mathbb{I} + m_z \sigma_z)/2$ & 
Bounded by $\delta / J = -m_z$ and $\delta /J =0$ &
$ \delta /J = -m_z$ &
Similar to the cases 1 and 3.
\\

\hline
6 &
$J \sigma_z$ & $\sqrt{\frac{\gamma}{2}}\sigma_x$ & $(\mathbb{I} + m_z \sigma_z)/2$ & 
Bounded by $\delta / J = -m_z$ and $\delta /J =0$ &
$ \delta /J = -m_z$ &
1. $K_3^{\mbox{opt}}$ possesses high values in the neighborhood of $\delta = - \mbox{tr}(\rho H)$ line.

2. The region of violation shrinks.
\\
\hline
 
\end{tabular}
\caption{Observations regarding the behaviors of $K_3^{\mbox{opt}}$ with the noise parameter, $\gamma$, for different dephasing operators and different set of initial states.}
\label{tab:obs}
\end{table*}

\subsection{Hamiltonian and dephasing operator are diagonal in complementary bases}

Let us now move to the case where the Hamiltonian and the dephasing operator are diagonal in a complementary bases i.e., 
\begin{eqnarray}
H=J \sigma_z, ~~~~~~~~~~~~~~~~~~~~ V=\sqrt{\frac{\gamma}{2}}\sigma_x.
\end{eqnarray}
In this case, even if the initial state is in a pure state, the expressions of $\Delta E$ and $K_3$ are  mathematically involved, and so we skip the expressions. From the expressions of $K_3$ and $\Delta E$, one can show that Theorem 2 also holds for such a noisy evolution, i.e., maximum value of $K_3^{\mbox{opt}}$ is attained for $\Delta E = - \mbox{tr}(\rho H)$ (see Fig. \ref{puregx} for pure states as the initial ones).

However, one can observe certain sharp contrast in the behavior compared to the previous case when $H$ and $V$ are in the same basis (see Fig. \ref{puregx} for three different values of decoherence parameter, $\gamma$, for the pure states given in Eq. (\ref{eq:pure_state})). Let us discuss the disparity observed in this scenario with the case in Sec. \ref{sec:same_basis} when the initial state is in a pure state.
\begin{enumerate}
\item The amount of violation of canonical LGI remains quite high even in the presence of sufficiently high amount of noise, especially near the $\delta = - \mbox{tr}(\rho H)$ line. Clearly, this feature indicates that the violation of LGI is more robust in presence of this kind of noise than the case when the Hamiltonian and the dephasing operator are in the same basis.

\item The region of violation in $(\alpha, \delta/J)$-plane decreases or shrinks with higher values of the decoherence parameter, which was not observed in the previous case.
\end{enumerate}
Let us now discuss whether these features are independent of the initial state which is the case for both noiseless and nosy scenarios discussed in preceding sections.
Let us consider an arbitrary mixed states, given in Eq. (\ref{eq:mixed_gen}) with
$m_y = m_z = 0$ and $m_x \neq 0$. We find that although the first observation for pure states remains qualitatively same, the violating region in the $(m_x, \delta / J)$-plane does not change with different values of $\gamma$. The feature remains unaltered if $m_x = m_z = 0$ and $m_y \neq 0$.

Let us now choose another set of mixed states with $m_x = m_y = 0$ and $m_z \neq 0$, i.e., 
$\rho = (\mathbb{I} + m_z \sigma_z)/2$,
with $m_z \in [0, 1]$. As shown in Fig. \ref{mixedzgx}, in this case, the close inspection of $K_3^{\mbox{opt}}$ reveals that the properties of the violation of LGI with the energy-constraint coincide with the case of the pure state. Specifically, the region of violation shrinks with the increase of $\gamma$. The analysis of $K_3^{\mbox{opt}}$ with the noise parameter is discussed in Table \ref{tab:obs} in a compact way.

\noindent \textbf{Remark:} Two classes of mixed states are chosen in such a way that magnetizations is non-vanishing either in the same basis  or in the complementary bases of the Hamiltonian.

We also consider initial states, where magnetizations in all three directions are non-zero. For such initial states, the behaviors of $K_3^{\mbox{opt}}$ with noise remain similar. However, the amount of decrement of the violation region increases with increasing $m_z$.

\subsection{Hamiltonian and dephasing operator are diagonal in bases that are at an angle of $\pi/4$ to each other}
We now move to a scenario which is intermediate to  the extremal cases considered in the previous two subsections, i.e., we now take $H$ and $V$ in such a way that they are diagonal in bases which are at an angle $\pi/4$ to each other. For this, we choose
\begin{eqnarray}
H=J \sigma_z, ~~~~~~~~~~~~~~~~~~~~ V=\frac{\sqrt{\gamma}}{2}(\sigma_x + \sigma_z).
\end{eqnarray}

From the continuity argument,  we can expect that for this kind of evolution the situation and the observations regarding the violation of canonical LGI will be somewhat in between what we have seen in previous two sections. In Fig. \ref{puregxz}, we plot the canonical LG expression, $K_3^{\mbox{opt}}$, for the pure states for different values of the decoherence parameters. Clearly, in this case, the region of violation shrinks with increasing values of $\gamma$. However, quite counter-intuitively, for this type of evolution, Theorem 2 does not hold, which can be seen from 
the figure as the maximum value of $K_3^{\mbox{opt}}$ shifts from the $\delta = - \mbox{tr}(\rho H)$ line. We numerically find that this shifting of maximum line of violation for canonical LGI occurs when $H$ and $V$ are not diagonal in the same or complementary bases.

\section{Conclusion}
\label{sec:conclusion}
Nonclassicality of temporal quantum correlations pertaining to non-commutative sequential measurements can be defined through violation of a  macrorealistic inequality, viz. the Leggett-Garg inequalities (LGI). These inequalities were invented to test the incompatibility of quantum mechanical systems with  macrorealism and
noninvasive measurability. Over the last few decades, such nonclassicality was shown to  be connected with various information processing tasks.

 Physical implementations of various information-theoretic protocols are performed within a relevant energy constraint
due to the limitations in laboratories or industries. For example, for infinite-dimensional systems, an energy constraint is crucial in quantum communication protocols. 

We investigated the optimal violation of a Leggett-Garg inequality under energy constraints, which we refer to as canonical LGI.
We proved that for an arbitrary initial state in two dimensions, an optimal violation of the ``$K_3$ inequality", an LGI, reaches its maximum value when the change in the energy is the negative of the average energy of the initial quantum state. 
We found that
this feature is true both in the noiseless scenario and in the presence of dephasing noise, 
when the Hamiltonian and the dephasing operator are diagonal in the same or in complementary bases, 
thereby showing the dependence of a violation of LGI under energy constraint on the initial states. This dpendance of the violation on the initial state is unlike that in the case of unconstrained energy.
Moreover we found  that when the Hamiltonian and the dephasing operator are not diagonal in the same basis, the violation of LGI near the maximum, changes slowly with the environmental noise, showing robustness of the phenomenon, although the  region of violation shrinks for both pure and for certain mixed states as initial ones. We observe that such shrinking of region of violation are uniform whenever the Hamiltonian and the dephasing operator are not diagonal in the same basis.

\acknowledgments 
This research was supported in part by the `INFOSYS scholarship for senior students'. The authors acknowledge computations performed at the cluster computing facility of Harish-Chandra Research Institute, Allahabad, India.


\begin{thebibliography}{99}






\bibitem{bell'65} J. S. Bell, Physics \textbf{1}, 195 (1965).
\bibitem{brunner'2014} N. Brunner, D. Cavalcanti, S. Pironio, V. Scarani, and S. Wehner, Rev. Mod. Phys. {\bf 86}, 419 (2014);
Rev. Mod. Phys. {\bf 86}, 839 (2014).

\bibitem{kochen} S. Kochen and E. P. Specker, J. Math. Mech. {\bf 17}, 59 (1967).
\bibitem{kcbs} A.A.Klyachko, M.A.Can, S.Binicioglu, and A. S. Shumovsky, Phys.Rev.Lett. {\bf 101}, 020403 (2008).
\bibitem{leggett'1985} A. J. Leggett and A. Garg, Phys. Rev. Lett. {\bf 54}, 857 (1985).

\bibitem{emary'2014} C. Emary, N. Lambert, and F. Nori, Rep. Prog. Phys. {\bf 77}, 016001 (2014). 

\bibitem{lgi_exp_sc} A. Palacios-Laloy, F. Mallet, F. Nguyen, P. Bertet, D. Vion, D. Esteve, and A. N. Korotkov,
Nat. Phys. 6, {\bf 442} (2010).


\bibitem{van'2000}
C. H. van der Wal, A. C. J. ter Haar, F. K. Wilhelm, R. N. Schouten, C. J. P. M. Harmans, T. P. Orlando, Seth Lloyd, and J. E. Mooij,
Science {\bf 290}, 773 (2000).

\bibitem{friedman'2000}
J. R. Friedman, V. Patel, W. Chen, S. K. Tolpygo and J. E. Lukens,
Nature {\bf 406}, 43 (2000).

\bibitem{roskov'2006}R. Roskov, A. M. Korotkov, and A. Mizel, Phys. Rev. Lett. {\bf 96}, 200404 (2006).
\bibitem{jordan'2006} A. N. Jordan, A. M. Korotkov, and M. Buttiker, Phys. Rev. Lett. {\bf 97}, 026805 (2006).
\bibitem{mahesh'2011} V. Athalye, S. S. Roy, and T. S. Mahesh, Phys. Rev. Lett. {\bf 107}, 130402 (2011).

\bibitem{leggett'2008} A. J. Leggett, Rep. Prog. Phys. {\bf 71}, 022001 (2008).


\bibitem{onofrio'1999} T. Calarco, M. Cini, and R. Onofrio, Europhys. Lett., {\bf 47}, 407 (1999).

\bibitem{leggett'2002} A. J. Leggett, J. Phys. Cond. Matt. {\bf 14}, R415 (2002).

\bibitem{kofler'2007}J. Kofler and {\v C}. Brukner, Phys. Rev. Lett. {\bf 99}, 180403 (2007).

\bibitem{avis'2010} D. Avis, P. Hayden, and M. M. Wilde, Phys. Rev. A {\bf 82}, 030102(R) (2010).



\bibitem{home'2013}D. Gangopadhyay, D. Home, and A. S. Roy, Phys. Rev. A {\bf 88}, 022115 (2013).

\bibitem{budroni'2013}C. Budroni, T. Moroder, M. Kleinmann, and O. G\"uhne; Phys. Rev. Lett. {\bf 111}, 020403 (2013).
\bibitem{budroni'2014}C. Budroni and C. Emary, Phys. Rev. Lett. {\bf 113}, 050401 (2014).
\bibitem{dressel'2014} J. Dressel and A. N. Korotkov, arXiv: 1310.6947 (2014).




\bibitem{mal'2016_1}S. Mal and A. S. Majumdar, Phys. Lett. A {\bf 380}, 2265 (2016).
\bibitem{maldas'2016} S. Mal, D. Das and D. Home, Phys. Rev. A {\bf 94}, 062117 (2016).


\bibitem{kofler'2013}J. Kofler and {\v C}. Brukner, Phys. Rev. A {\bf 87}, 052115 (2013).

\bibitem{udevi'2013} A. R. U. Devi, H. S. Karthik, Sudha, and A. K. Rajagopal, Phys. Rev. A {\bf 87}, 052103 (2013).

\bibitem{mal'2015} D. Saha, S. Mal, P. K. Panigrahi, and D. Home, Phys. Rev. A {\bf 91}, 032117 (2015).





\bibitem{cabello} A. Cabello, V. D'Ambrosio, E. Nagali, and F. Sciarrino, Phys. rev. A {\bf 84} 030302 (2011).
\bibitem{arvind} J. Singh, K. Bharti, and Arvind, Phys. Rev. A {\bf 95}, 062333 (2017).
\bibitem{shane} S. Mansfield, and E. Kashefi, arXiv: 1801.08150 (2018).




\bibitem{brukner'2004} {\v C}. Brukner, S. Taylor, S. Cheung, V. Vedral, arXiv:quant-ph/0402127 (2004).

\bibitem{mal'2016_2} S. Mal, M. Banik, and S. K. Choudhury, Quant. Inf. Process. {\bf 15}, 2993 (2016). 

\bibitem{shenoy'2017} A. Shenoy, H. S. Aravinda, R. Srikanth, and D. Home, Phys. Lett. A {\bf 381}, 2478 (2017).




\bibitem{holevo'1973} A. S. Holevo, Probl. Peredachi Inf. {\bf 9}, 3 (1973).

\bibitem{schumacher'1996} B. Schumacher, M.  Westmoreland, and W.  K. Wootters, Phys.  Rev.  Lett. {\bf 76}, 3452 (1996).

\bibitem{lloyd'2003}V. Giovannetti, S. Lloyd, L. Maccone, and P. W. Shor, Phys. Rev. Lett. {\bf 91} 047901 (2003).

\bibitem{asen'2005} A. Sen(De), U. Sen, B. Gromek, D. Bru{\ss}, and M. Lewenstein, Phys. Rev. Lett. {\bf 95}, 260503 (2005).
\bibitem{asen'2007}A. Sen(De), U. Sen, B. Gromek, D. Bru{\ss},and  M. Lewenstein, Phys. Rev. A {\bf 75} 022331 (2007).
\bibitem{eisert'2002} J. Eisert, Ch. Simon, and M. B. Plenio, J. Phys. A {\bf 35}, 3911 (2002).


\bibitem{shirokov'2004} M. E. Shirokov, Theory of probability and its Applications {\bf 52}, 250 (2007).

\bibitem{thcost1} M. Huber, M. Perarnau-Llobet, K. V. Hovhannisyan, P. Skrzypczyk, C. Kl\"ockl, N. Brunner and A. Ac\'in, New J. Phys. {\bf 17}, 065008(2015).
\bibitem{thcos} D. E. Bruschi, M. Perarnau-Llobet, N. Friis, K. V. Hovhannisyan, and M. Huber, Phys. Rev. E {\bf 91}, 032118 (2015).


\bibitem{horo'1997} P. Horodecki, Phys. Lett. A {\bf 232}, 333 (1997).
\bibitem{horo'1998} M. Horodecki, P. Horodecki, and R. Horodecki, Phys. Rev. Lett. {\bf 80}, 5239 (1998).
\bibitem{horo'98} P. Horodecki, R. Horodecki, and M. Horodecki, Acta Phys. Slov. {\bf 48}, 141 (1998).
\bibitem{horo'01} R. Horodecki, M. Horodecki, and P. Horodecki, Phys. Rev. A {\bf 63}, 022310 (2001).

\bibitem{tamoghna'2017} T. Das, A. Kumar, A. K. Pal, N. Shukla, A. Sen(De), and U. Sen, Phys. Lett. A {\bf 381}, 3529 (2017).


\end{thebibliography}
\end{document}